\documentstyle[12pt,aasms]{article}

\def\fun#1#2{\lower0.837ex\vbox{\baselineskip0ex\lineskip0.209ex
  \ialign{$\mathsurround=0ex#1\hfil##\hfil$\crcr#2\crcr\sim\crcr}}}

\def\msun{M_\odot}

\def\sles{\lower2pt\hbox{$\buildrel {\scriptstyle <}
   \over {\scriptstyle\sim}$}}
 
\def\sgreat{\lower2pt\hbox{$\buildrel {\scriptstyle >}
   \over {\scriptstyle\sim}$}}

\begin{document}

\title{ The Nature of the Giant Outbursts in 
the Bursting Pulsar GRO J1744-28 }

\author{  John K. Cannizzo$^1$}
\affil{e-mail: cannizzo@lheavx.gsfc.nasa.gov}
\affil{Goddard Space Flight Center}
\affil{NASA/GSFC/Laboratory for High Energy Astrophysics, 
Code 662, Greenbelt, MD 20771}
\authoraddr{NASA/GSFC/Laboratory for High Energy Astrophysics, 
Code 662, Greenbelt, MD 20771}

\vskip 1truein
\ {\it }$^1$ Universities Space Research Association

\begin{abstract}

We investigate the 
possible role of an accretion disk
instability in producing the
giant outbursts
seen in GRO J1744-28.
Specifically,
we study the global,
time dependent
evolution  of the Lightman-Eardley
instability
which can develop 
near the inner edge of 
an accretion disk
when the radiation pressure
becomes comparable to the
gas pressure.
Broadly speaking, our results are
compatible with earlier works
by Taam \& Lin and by Lasota \& Pelat.
The uniqueness of GRO J1744-28
appears to be associated with the constraint
that, 
in order for outbursts to occur,
 the rate of accretion
at the inner edge must be
within a  narrow range
just above
the critical accretion rate
at which radiation pressure is beginning
to become significant.

\medskip
\medskip

Subject Headings: accretion disks: instabilities $-$  X-rays: bursts $-$ 
                      pulsars: individual (GRO J1744-28)

\end{abstract}

\section{ Background }

 On 2 December  1995 the Burst and Transient
Source Experiment (BATSE)
on  the {\it
         Compton Gamma Ray Observatory}
({\it CGRO})
discovered a new source
which maintained a high 
persistent flux level,
and also showed outbursts
with a duration of about 10 seconds
during which time the
X-ray flux increased
by about 4-5 (Kouveliotou et al. 1996).
The source has subsequently been studied
in detail by the {\it  Rossi X-ray Timing
Explorer} ($=${\it RXTE}, 
  cf. Swank et al. 1996).
Subsequent
 investigation of this
unusual source showed
it to reside in a binary star
system
with an orbital period of 11.8 days (Finger et al. 1996).
The pulsar has a spin period of 0.467 seconds (Finger et al. 1996) 
and concomitant 
 corotation
radius $1\times10^8$ cm $(M_1/1.4\msun)^{1/3}$.
Several papers
have already appeared which
attempt to address
broad issues connected
with the evolutionary status
of the system
and the strength of the magnetic
field on the pulsar 
(e.g., Lewin et al. 1996, Daumerie et al. 1996, Lamb et al. 1996,
Sturner \& Dermer 1996).

The question has naturally arisen 
as to the cause of the outbursts.
The two main models being discussed
at present are 
(1) thermonuclear flashes of
material accumulated
on the surface of the 
neutron star,
           and
(2) some as yet unspecified
accretion disk instability which causes
a periodic storage and dumping of some
material in the inner disk.
In the light curve 
one sees a dip and recovery
period following each outburst,
during which time the light asymptotically
approaches its pre-outburst level.
Also, the spectrum does not change
dramatically going from quiescence 
to outburst.
This behavior 
may not favor the thermonuclear flash model
(Lewin et al. 1993, 1996).

The most obvious disk instability
to invoke for the outbursts
is the Lightman-Eardley (LE) instability
(Lightman \& Eardley 1974).
This instability develops
in viscous accretion disks when the
radiation pressure becomes 
comparable  to the gas pressure.
Several papers have already investigated
the global, nonlinear evolution
of   accretion disks which are
LE unstable (Taam \& Lin 1984=TL, Lasota \& Pelat 1991=LP).
These investigators
showed that
the instability first develops at the
inner edge of the disk
and propagates outward as a heating wave
of high viscosity material,
much like the heating wave
associated with the classical
limit cycle instability in dwarf novae and X-ray novae.
The instability does not propagate far,
and the heated matter with its increased
viscosity rapidly accretes onto the central star
and produces a brief burst.

Figure 1 shows a $\sim1$ hr stretch of data obtained
13 March 1996 with the Proportional Counter Array (PCA)
instrument on {\it RXTE}.
 Three outbursts were seen $-$ with durations of
$\sim$10 seconds  and     recurrence times of $\sim$1000 seconds.
The light curves shown in TL and LP bear a striking
resemblance to those seen in GRO J1744-28,
in particular one sees the broad dip
and recovery phase after a burst has ended.
The main shortcoming of the TL and LP bursts
insofar as they might pertain to GRO J1744-28
is that they recur every few seconds and
last for less than one second. These time scales are
    both much faster than observed.
In
 their modeling TL and LP  
(1) set the viscosity parameter $\alpha$
to its maximal value of 1,
and they 
(2) took the inner
disk edge to equal roughly
the radius of the neutron star ($\sim10^6$ cm),
whereas we suspect that in
GRO J1744-28 
the pulsar
has a strong magnetic
field. 
In this {\it Letter} 
we quantify the increases in the outburst
time scales due to setting $\alpha$
to a more reasonable value,
and to increasing $r_{\rm inner}$.

\section {General Physical Considerations}

The limit cycle found by TL and LP is
unlike the standard  limit cycle
for dwarf novae (see Cannizzo 1993a for 
a recent review).
In the standard dwarf nova  model,
the locus of steady state
solutions forms an S-curve
when plotted as effective
temperature
versus surface density (Meyer \& Meyer-Hofmeister 1981).
For the LE instability, however,
there is 
    no upper stable branch.
(This situation has recently changed
  [Abramowicz et al. 1988].)
Following the discovery of the LE
instability in the 1970's,
this lack of an upper stable branch
was viewed as a
severe limitation of
the instability which might
restrict its 
 usefulness.
The earliest time dependent studies
of the LE instability
solved 
only the viscous diffusion equation
for surface density.
TL and LP
simultaneously solved
both the diffusion equation
and the thermal
 energy equation for temperature,
and therefore considered
rather more general 
accretion  disks which need be neither
steady (i.e., ${\dot M} $ constant
with radius) nor in thermal  equilibrium.

The fundamental obstacle
to progress in accretion disk
research has been the lack
of understanding of the
physical mechanism responsible
for the viscous dissipation.
In a recent study of the
time dependent evolution
of accretion disks
in black hole X-ray binaries,
Cannizzo et al. (1995$=$CCL)
found that, to reproduce the observed
$\sim30-40$ day exponential
decays of the X-ray fluxes
in the 
soft X-ray transients,
the  Shakura-Sunyaev viscosity parameter 
$\alpha$ must take the form
$\alpha=\alpha_0(h/r)^n$,
where $\alpha_0 \simeq 50$ and $n=1.5$.
In this study we adopt  $\alpha=50(h/r)^{1.5}$,
keeping in mind that this form may not be valid
when radiation pressure begins to play a role.
In the limit of large $\alpha$, we do not allow
$\alpha$ to exceed $0.25$.

One may gain some rough
understanding of the criterion
for instability by
considering scalings
for physical properties
at the
local maximum in $\Sigma$
associated with the transition to radiation
pressure domination.
Using    the Shakura-Sunyaev
  ``middle'' region for which $P=P_{\rm gas}$
and $\kappa=\kappa_{\rm es} = 0.34$ cm$^2$ g$^{-1}$,
     the rate of accretion at which the
     gas and radiation  pressure are equal is
 \begin{equation}
{\dot M}_{\rm crit} = 8.3\times 10^{18}  \ {\rm g} \  {\rm  s}^{-1} \
\alpha^{-1/8} \Omega^{-7/8} \mu^{-1/2}
\end{equation}
\noindent 
(Shakura \& Sunyaev 1973).
Here
$\Omega$ is the local angular frequency
(assumed to be Keplerian)
and $\mu$ is the mean molecular weight ($=0.617$).
The
 temperature
associated with the LE unstable branch
and with $\Sigma_{\rm crit}$
is 
$T_{\rm crit} = 2.11\times 10^6 \ {\rm K} \ \alpha^{-1/4} \ \Omega^{1/4}$.
Combining this
with the law
$\alpha=50(h/r)^{1.5}$
and 
the condition of hydrostatic equilibrium
$h\Omega = \sqrt{{\cal R}T/\mu}$
gives
\begin{equation}
{\dot M}_{\rm crit} = 1.5\times 10^{18} \ {\rm g } \ {\rm s}^{-1} \ 
       \ r_8^{1.26} \ m_1^{-0.55}.
\end{equation}
The viscous diffusion time at the inner edge 
$
t_{\nu, \ {\rm crit} } 
=$
$ {r_{\rm inner}}^2 / \nu_{\rm crit} = 1200$   s
     $ \ r_8^{0.58} \ m_1^{ 0.79}$,
where 
$\nu_{\rm crit} =$ 
$ (2\alpha_{\rm crit}/3\Omega_{\rm inner})$
$ ({\cal R}T_{\rm crit}/\mu)$,
$r_8=r_{\rm inner}/10^8$ cm,
and $m_1 = M_1/1\msun$.
This time scale is roughly the observed
recurrence time scale for outbursts in GRO J1744-28.

\section {Model}

The model we
use is 
the same as that
described in detail in TL and LP.
This is a time dependent model
which follows the evolution of surface density
and temperature. 
The heating and cooling functions are evaluated
separately, and the radial flow of energy $-$
from both advection and diffusion $-$
is included. The  radial pressure gradient
and   departures from Keplerian flow are not included.
Our numerical code
is a modified version of the
one used previously for
modeling dwarf novae and X-ray novae
(cf. Cannizzo 1993b,  Cannizzo et al. 1995).
In this study we utilize a grid spacing
for which $\Delta r\propto \sqrt{r}$,
and take $N=40$ radial grid points.
For the pulsar mass we adopt $M_1 = 1.4\msun$.
The inner and outer disk radii are taken to be
$r_{\rm inner}= 10^{7.5}$ cm and   
$r_{\rm outer} =  10^9$ cm, respectively.
The value of $r_{\rm inner}$
was chosen to be less than the corotation radius,
in order to be consistent with the observed
spin-up of the pulsar.
The value of $r_{\rm outer}$
was chosen to be large enough
so that the heating
front never comes close to reaching it.

Figure 2 shows a sample light curve
from our model, for a mass transfer
rate into the outer disk 
of $1.5\times 10^{18}$ g s$^{-1}$.
We show a 3000 s history of
(1) the rate of mass loss at the inner disk edge,
and 
(2) the mass of the accretion disk.
For this model, the observed  burst durations of $\sim10$ s
and recurrence times of $\sim1000$ s are reproduced.
There is also a dip and recovery following
  each major  outburst $-$
although the dip is 
         somewhat deeper
and faster than observed.

Figure 3 shows the evolution of the
disk in $Z=P_{\rm radiation}/P_{\rm gas}$,
$\Sigma$, and $h/r$.
We show 100 seconds of evolution centered on the 
second burst in Figure 2, spanning the time
from 1480 to 1580 seconds.
Each curve in Fig. 3 is separated by 2 seconds.
The $t=0$ curve corresponds to
just prior to the onset of the outburst.
The ratio $Z$ has just begun to exceed $\sim1$
in the inner disk.
At slightly later times
$Z$ increases rapidly to $\sim10$ as the local
gas heats and matter accretes onto the pulsar,
but then quickly drops back down to $<1$
as the decreased surface density
forces a return to the gas pressure dominated branch.
The second panel in Figure 3
reveals how the heating front
is propagated as a local enhancement in $\Sigma$.
This is quite similar to what is seen
in computations of the classical limit cycle instability
(Cannizzo 1993b).
Finally, the third panel shows that $h/r$
is always considerably less than unity $-$
varying from a few percent in quiescence to
$\sim0.1$ in outburst.

Figure 4 shows the evolution of the disk at two
radial grid points near the inner disk
edge in $(T,\Sigma)$ space.
The dashed line shows the standard 
equilibrium relation derived from
taking the heating and cooling functions to be equal
and assuming the viscosity to 
couple to $P_{\rm gas} + P_{\rm radiation}$
(as in our model).
The dotted curve shows the equilibrium track for which
the viscosity couples to the gas pressure only.
In quiescence the evolutionary track closely follows
the thermal equilibrium curve.
The deviation
between the actual evolution and the equilibrium
curve 
    becomes stronger for the
evolution at $4\times 10^7$ cm
as $\Sigma_{\rm crit}$ is approached.
In Figure 4  we see that, after an annulus
has made a transition to the LE state,
it eventually migrates to smaller surface density
(due to removal of matter from the inner edge),
and finally proceeds back down to the gas
pressure dominated branch. At this point
we observe  the build-up
in $\Sigma$ associated with material from further out
flowing in to refill the cavity.
The evolution for the annulus at $1.1\times 10^8$ cm
is slightly different.
This annulus is further away from the site
of the initial trigger, and only makes the transition
to the high state by virtue of 
having the heating front sweep past.
The surface density
drops after the heating wave passes, and
rapid cooling forces a return
to the gas pressure dominated branch.
The evolutionary trends shown in Figure 4
are similar to those displayed by LP.

\section{Discussion and Conclusions}

We have shown how the LE instability
operating in the inner edge of an accretion
disk can be used to account for the giant outbursts
seen in the bursting pulsar GRO J1744-28.
In accordance with the time dependent
calculations of TL and LP,
we find that
the LE instability never has a chance to go
strongly into the non-linear regime
because
the rapid loss of material from the inner
edge which happens once the inner disk
starts to become LE unstable prevents
$P_{\rm radiation}/P_{\rm gas}$
from running away to a high value.
The material that is evacuated onto
the pulsar during an accretion event is replenished
by material  flowing in from
    further out,
hence the dip and recovery in the
light curve following an outburst.
TL and LP found much shorter outburst
time scales 
(i.e., $t_{\rm recurrence} \simeq $ a few seconds)
than we do 
because they took larger $\alpha$ values ($\alpha=1$)
and smaller $r_{\rm inner}$ values ($\simeq 10^6$ cm).

The uniqueness of GRO J1744-28 might
be associated with the possibility
that
${\dot M}$
onto the pulsar could 
         only exceed
${\dot M}_{\rm crit}(r_{\rm inner})$
by some marginal amount $-$ whose precise
value remains to be determined $-$
for the outbursts to occur.
If the mass transfer rate were too large,
there may exist a permanent region
of radiation pressure 
domination at smaller radii, 
and not the oscillation we find between 
$P_{\rm radiation}>P_{\rm gas}$ and
$P_{\rm radiation}<P_{\rm gas}$. 
For most low mass X-ray binaries
it is probably the case that,
lacking a strongly magnetized pulsar,
the inner disk edge is close to the neutron star
or black hole. 
 The fact that persistent LMXBs
such as Cyg X-2 and Sco X-1 are seen at all must
mean that they have accretion rates
comparable to GRO J1744-28.
Therefore, since
${\dot M}_{\rm crit}(r_{\rm inner})$
scales with $r_{\rm inner}$, one would
expect
$ {\dot M}/{\dot M}_{\rm crit}(r_{\rm inner})>>1$
for these systems,
and outbursts of the type considered
in this {\it Letter} could not occur.
Further work must be done to determine how
the size of the zone of instability varies with 
$r_{\rm inner}$ and hence
${\dot M}_{\rm crit}$.

One consequence of eqn. (2) is that, as the
long-term, mean
  mass accretion rate onto the pulsar decreases
and the  persistent
flux diminishes,
the outbursts in GRO J1744-28 
     should cease. 
As of  $\sim1$ May 1996, plus or minus several days,
    with the persistent {\it RXTE}
PCA flux level at $\sim0.1$ Cb,  the giant outbursts
have indeed stopped occurring.

\acknowledgments

It is a pleasure to
acknowledge useful criticisms
from Charles Dermer.
We also thank
Jean-Pierre Lasota
and Didier Pelat
at the Meudon Observatory
in Paris
for useful discussions
during a visit,
and
 Laurence Taff,
Alex Storrs,
and Ben Zellner at the Space Telescope Science
Institute
for allowing us generous use
of CPU time on
 their DEC AXP
workstations.
JKC
 was supported 
through the visiting
scientist program under
 the Universities Space Research Association
(USRA contract NAS5-32484)
in the Laboratory for High Energy Astrophysics
at Goddard Space Flight Center.

\vfil\eject
\centerline{ FIGURE CAPTIONS }
\medskip

Figure 1. A $\sim1$ hr light curve in X-rays of GRO J1744-28
    taken 13 March 1996 by the PCA instrument on {\it  RXTE}.
       The persistent flux
from the source was slightly less than
1 Crab ($=2.4\times10^{-8}$ erg cm$^{-2}$ s$^{-1}$).
The time origin corresponds to JD 2450155.8789.

Figure 2.  A sample 3000 second run from our model
showing (a) the rate of mass loss from the
inner disk onto the pulsar (which provides
the accretion luminosity) and (b) the mass of the disk.
The background accretion level
into the outer disk  is $1.5\times 10^{18}$ g s$^{-1}$.

Figure 3. The evolution of
     $Z=P_{\rm radiation}/P_{\rm gas}$, 
    surface density  $\Sigma$, and
local aspect ratio $h/r$.
The development shown spans the times $t=1480$ s to $t=1580$ s
in Fig. 2.
The dashed curves show the initial configuration
of the disk, just before the outburst is triggered
near the inner edge. The spacing between each curve is 2 s.

Figure 4. The evolution of two annuli in $(T,\Sigma)$ space
for one complete cycle.
The two panels are for
(a) $r=4.0\times 10^7$ cm, and
(b) $r=1.1\times 10^8$ cm.
The
   dashed and dotted curves give  the equilibrium relations
obtained by setting the heating and cooling functions
equal: $\Sigma^2 = (32acT^4)/(27\kappa\nu\Omega^2)$.
The dashed curves are  for $\nu$ scaling as $P_g + P_r$
as was used in the model,
while the dotted curves are  for $\nu$ scaling as $P_g$.

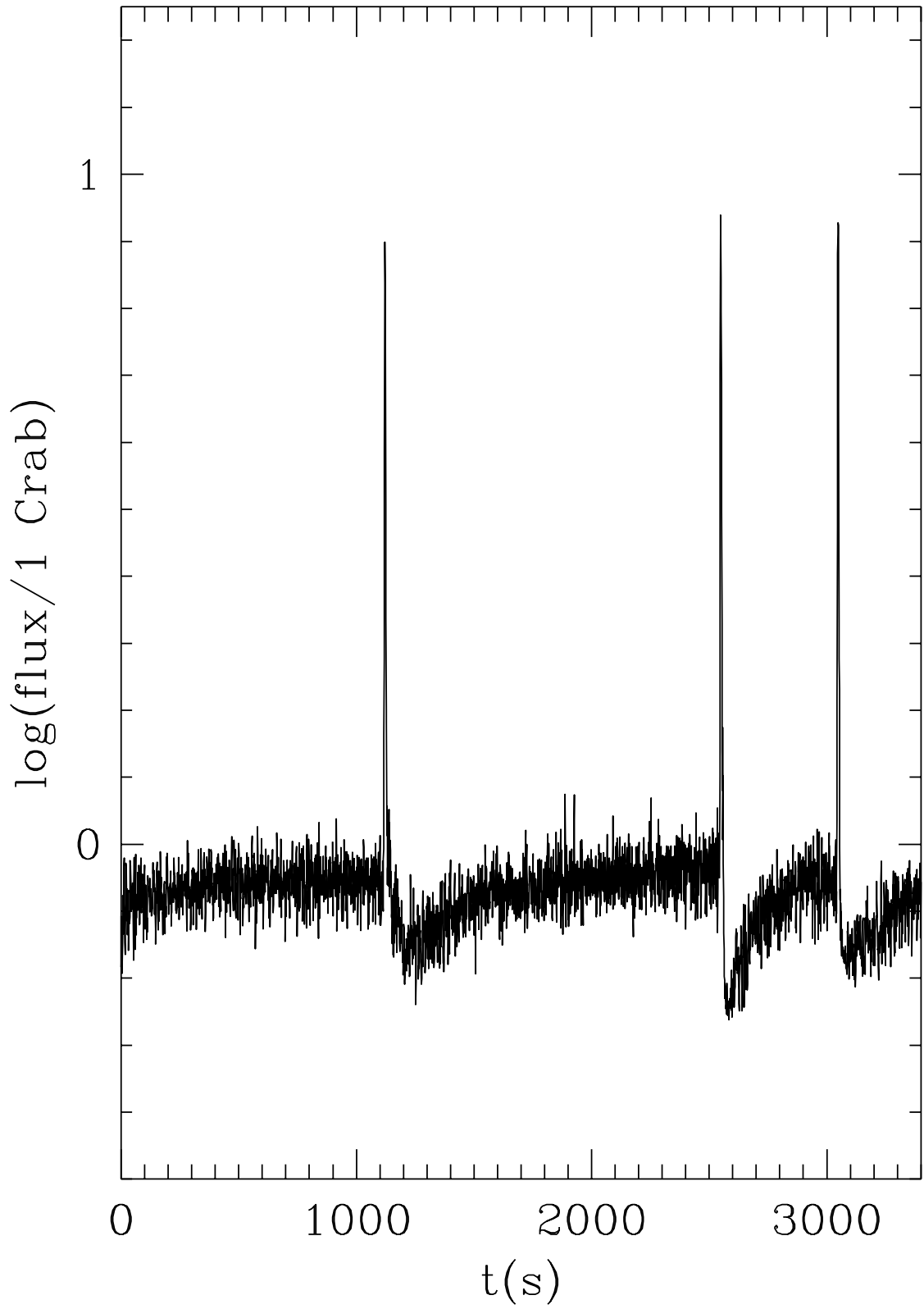
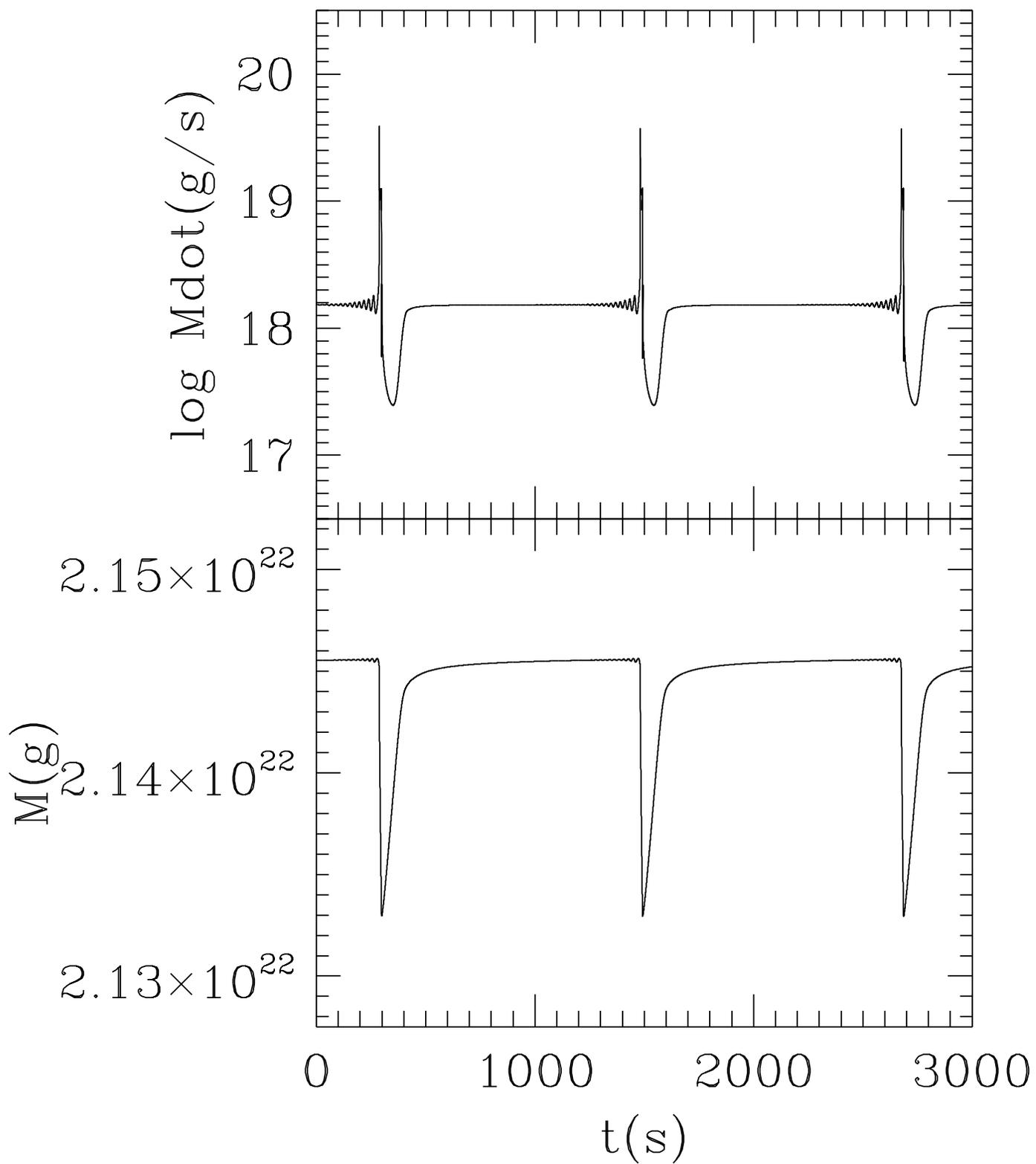
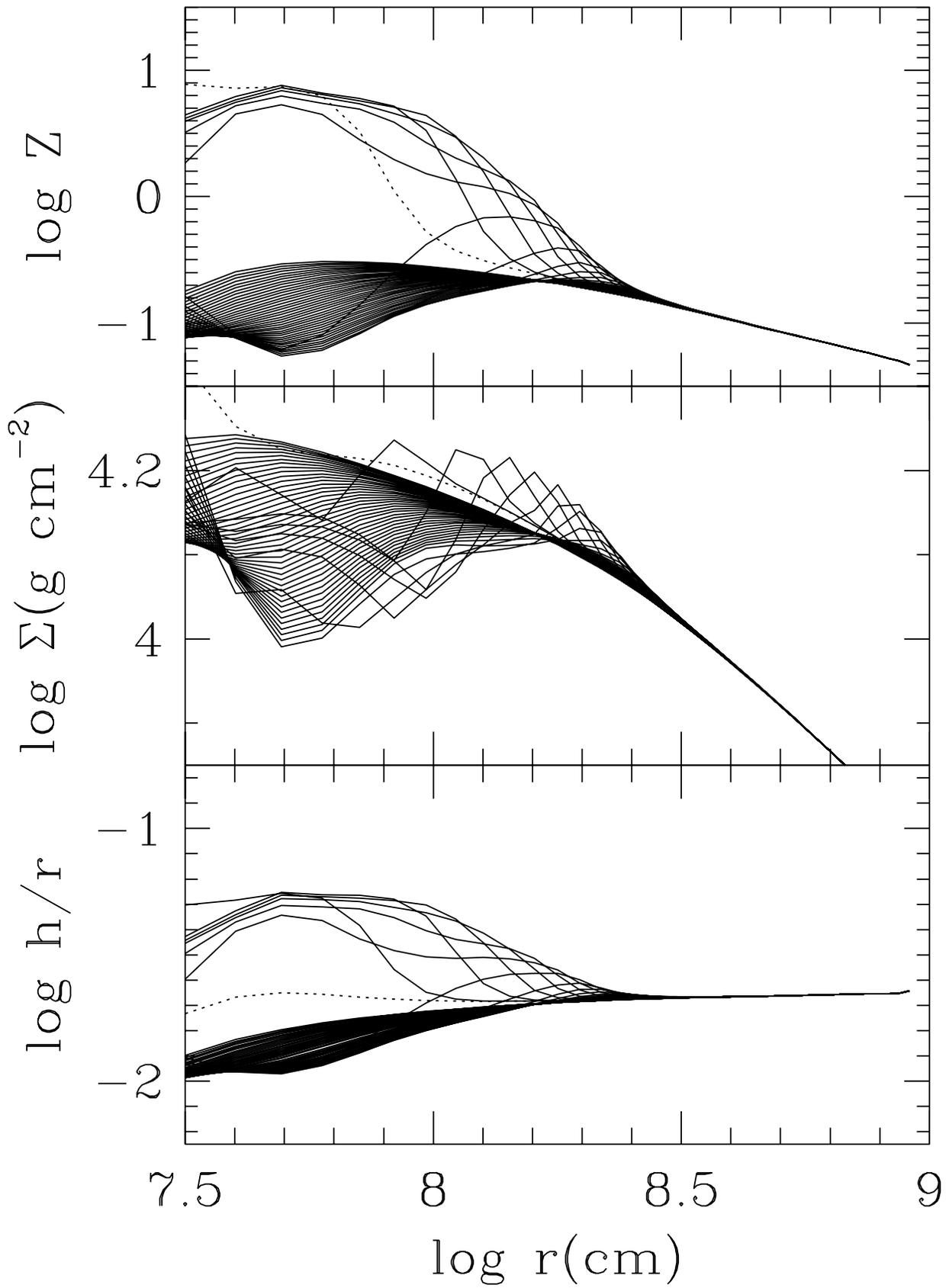
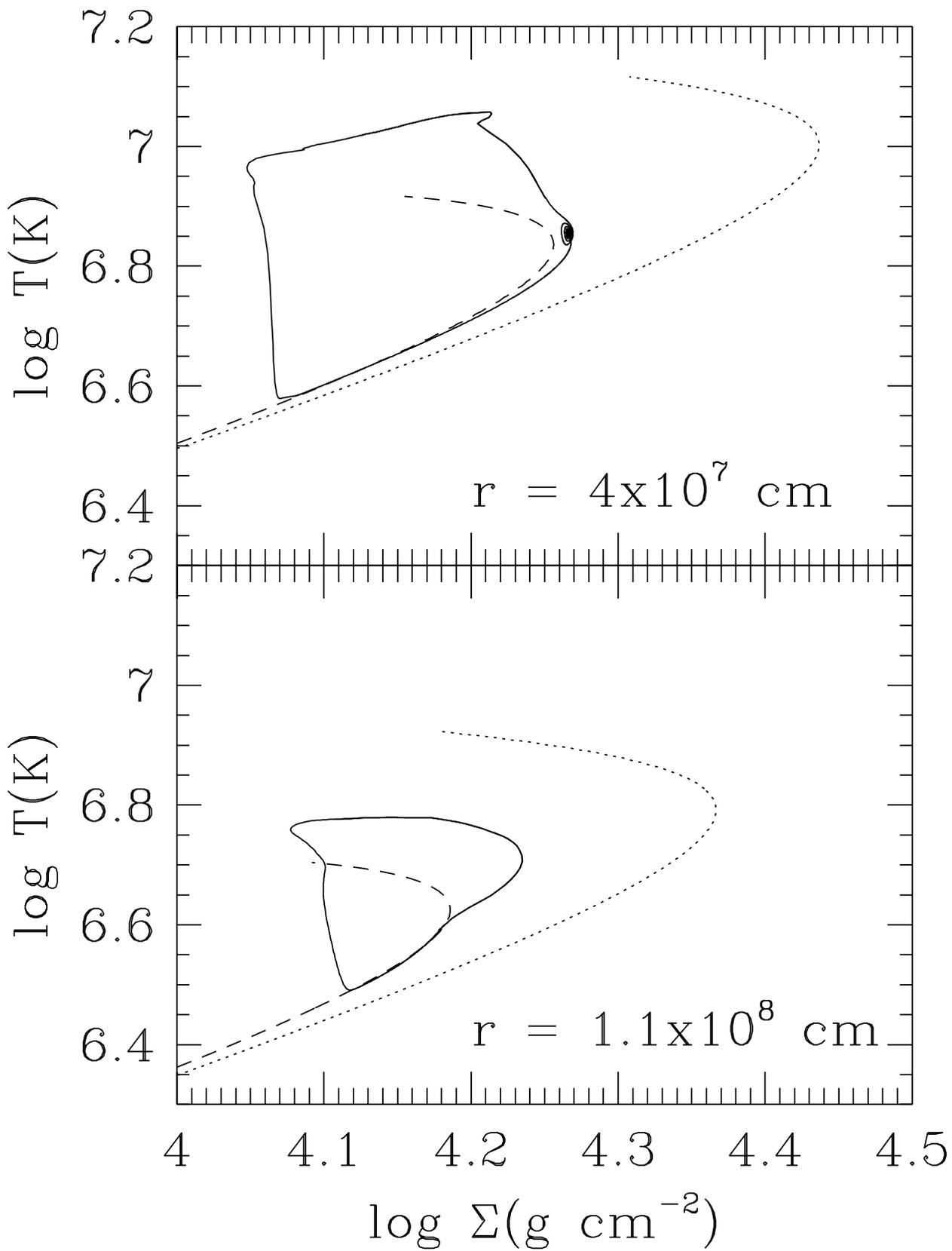
\end{document}